\documentclass{article}
\usepackage{hiph-art}
\volnumber{19} \issuenumber{1} \edyear{2004}                             
\frompage{000} \topage{000}                                              
\recrevdate{23 November 2005}                                              
\def\rightmark{Thermalization and RHIC Data}
\def\leftmark{G.Odyniec}
\usepackage{psfig}

\title{Signs of Thermalization from RHIC Experiments}
\authors{
{Grazyna Odyniec %
\index{Odyniec, G.} 
}\\[2.812mm]
{\normalsize
\hspace*{-8pt} Lawrence Berkeley National Laboratory\\
Berkeley, CA 94720, USA\\[0.2ex]
}}

\abstract{Selected results from the first five years of RHIC data
taking are reviewed with emphasis on evidence for thermalization in
central Au+Au collisions at $\sqrt{s_{NN}}=200$ GeV.}
\keyword{thermalization, elliptic flow, quark coalescence, jets}

\PACS{24.85.+p, 25.75.Nq, 25.75.-q}

\makeindex
\begin{document}

\maketitle

\section{Introduction}\label{intro}
The U.S. Relativistic Heavy Ion Collider (RHIC) was built primarily
to create and study the thermalized system of deconfined quarks and
gluons, called the Quark Gluon Plasma (QGP), predicted by QCD at
high energy densities. The first five, very successful, years of
RHIC operation provided scientists with an enormous wealth of data
leading, subsequently, to the conclusion that a new form of hot and
dense nuclear matter was created in Au+Au collisions at
$\sqrt{s_{NN}}=200$ GeV with energy density significantly exceeding
QCD estimates of critical energy density for a hadron gas - QGP
phase transition. The properties of this new form of nuclear matter,
however, are as yet, far from being known and understood.

A number of quite unexpected observations have been reported.
Perhaps the most striking is evidence for partonic collectivity and
jet quenching, both related to thermalization, discussed broadly at
this conference \cite{Vi}.

Note that some measurements are either preliminary or statistically
limited, and therefore their interpretation ought to be considered
as tentative at the present time.

\section{The Early Stage of the Collision - Partonic Collectivity and Thermalization}\label{techno}
One of the first and the most surprising RHIC results was an
elliptic flow measurement in Au+Au collisions at $\sqrt{s_{NN}}=$
130 and 200 GeV. This is a particularly important observable which
provides information on the matter created very early in the
collision. During the collision the initial coordinate-space
anisotropy of the system (the collision overlap region is elliptic
in shape in non-central nucleus-nucleus events) is converted by
secondary interactions and density gradients, built up in the
collision center, into an anisotropy in the final momentum-space.
This is commonly parametrized by a Fourier expansion series.
Elliptic flow, $v_2$, is defined as the second harmonic coefficient
of the azimuthal asymmetry with respect to the reaction plane (the
plane defined by the beam and impact parameter directions). Of
course, the efficiency of this conversion depends on the medium
properties. Note that elliptic flow possesses self-quenching
properties - once the spatial anisotropy disappears during
expansion, the development of elliptic flow stops. Therefore it is
primarily sensitive to the early stage Equation-Of-State
(EOS)\cite{TH6,TH7,TH8}

\begin{figure}[htb]

                 \insertplot{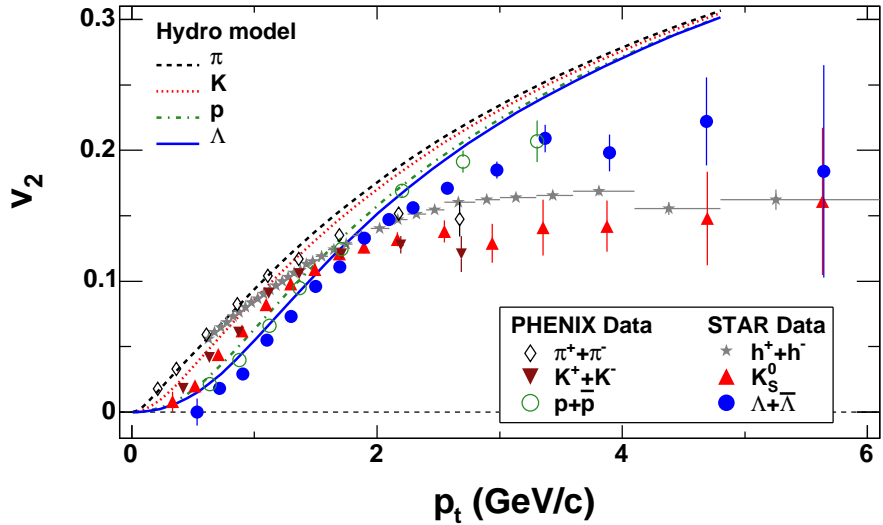}
\caption[]{$v_2$ for variety of particles from a minimum-bias sample
of Au+Au collisions at $\sqrt{s_{NN}}=$ 200 GeV measued by the STAR
\cite{ST1} and PHENIX \cite{PH1} collaborations. Curves show the
results from hydrodynamical model calculations \cite{TH1}
\label{fig1}}
\end{figure}

Figure 1 shows the elliptic flow parameter $v_2$ for pions, kaons,
protons, and lambda hyperons measured by two major RHIC experiments,
STAR \cite{ST1} and PHENIX \cite{PH1}. A large $v_2$ is observed for
all particle species indicating interactions at the early stage. For
$p_t$ $<$ 1.5 GeV/c, $v_2$ increases gradually with $p_t$. This
trend is well described by hydrodynamical model calculations
\cite{TH1}. More interestingly, the mass ordering, characteristic of
a common velocity field, with heavier particles exhibiting lower
values of $v_2$, is in good agreement with hydrodynamical models.
This indicates the presence of some (perhaps significant) degree of
thermalization at the early stage of the collision. At higher $p_t$
hydrodynamical calculations break down, as expected. The model
over-predicts the measured $v_2$ values and the particle-type
dependence is reversed. Above $p_t$ of about 2 GeV/c the shape of
the $v_2$ distribution flattens out (saturation) while mesons and
baryons form two distinct bands. The question of how $v_2$ was
established at these $p_t$ (above 2 GeV/c) remains open.

The run IV, data taken in 2004 \cite{ST2,PH2,ST3}, with very high
statistics and greater coverage for identified particles, extends
observations to even higher $p_t$ (up to 9-10 GeV/c). And, indeed,
even at the highest measured $p_t$, a large value of $v_2$ is still
present - see Figure 2 \cite{ST4}. The error bars in Figure 2 show
the statistical uncertainty, while the systematic errors are
presented as bands.

\begin{figure}[htb]

\insertplot{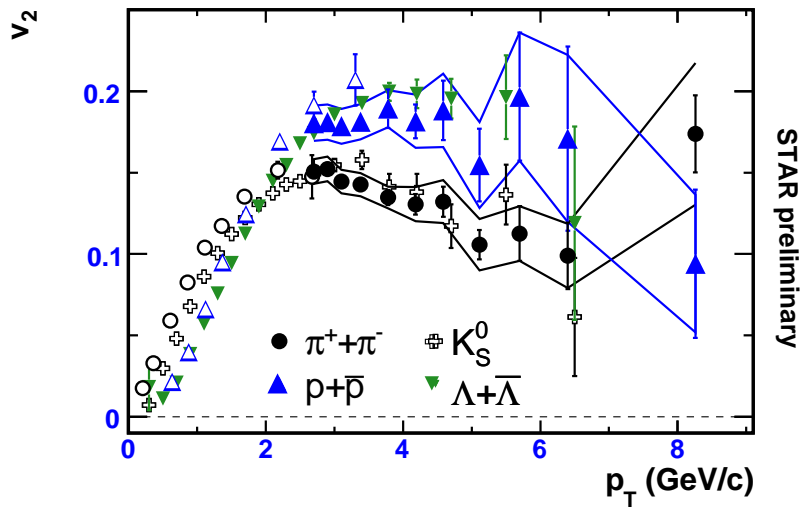}
 \caption[]{Azimuthal anisotropy $v_2$ for $\pi^+$+$\pi^-$ and $p$+$\bar{p}$
 in 200 GeV minimum bias Au+Au collisions. $K_0$ and $\Lambda$
 +$\bar{\Lambda}$ $v_2$ are shown for comparison. Open symbols
for $\pi^+$+$\pi^-$ and $p$+$\bar{p}$ were taken from \cite{PH1}.}
\label{fig2}
\end{figure}

\begin{figure}[htb]

\insertplot{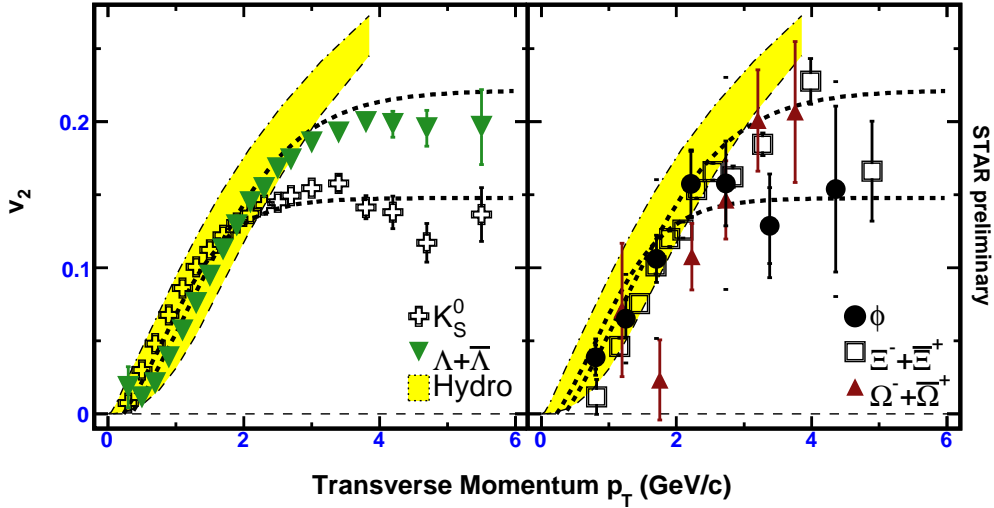} \caption[]{Azimuthal anisotropy $v_2$ for
strange hadrons (left) and multi-strange hadrons (right) in 200 GeV
minimum bias Au+Au collisions. The dash lines show a common fit to
the $K^0$ and $\Lambda$+$\bar{\Lambda}$ data \cite{Do}. Hydrodynamic
model calculations are shown as shaded areas \cite{Hu}.}
\label{fig3}
\end{figure}

The run IV $v_2$ measurements, with unprecedented accuracy for
multi-strange hadrons ($\Xi$, $\Omega$ and $\phi$), show large $v_2$
values and similar grouping to those observed for non-strange
particles as shown in Figure 3 \cite{ST4,ST5}. The independence of
elliptic flow on hadronic cross sections  ($\Xi$, $\Omega$ and
$\phi$ have very small hadronic cross-sections compared to
non-strange particles) suggests that $v_2$ was developed in the
partonic stage very early in the collision, before the hadronization
process took place. For comparison, Figure 3 shows the range of
$v_2$ from hydrodynamical calculations. The saturation value of
$v_2$ above $p_t$ of 2 GeV/c for mesons is about 2/3 of that for
baryons. This pattern, which holds for $\pi, p, K,\Lambda, \Xi$ and
with larger error bars also for $\phi$ and $\Omega$, indicates that
$v_2$ distributions can be scaled by the number of constituent
quarks ($n_q$) in the hadrons under study (i.e. $n_q$=2 for $\phi$,
$n_q$=3 for $\Omega$ etc). This observation points towards relevance
of constituent quark degrees-of-freedom. Figure 4, shows $v_2$ as a
function of $p_t$ where $v_2$ and $p_t$ are both scaled by the
number of constituent quarks. All identified particles fall on one
curve supporting the picture of hadronisation by coalescence or
recombination of constituent quarks \cite{TH2,TH3,TH4,TH5}. A
polynomial function was fitted to the scaled values. The bottom
panel shows the ratio between the measurements and the fit. At low
$p_t/n$ ($<$ 0.75 GeV/c) the observed deviations from the fit follow
a mass-ordering which is expected from hydrodynamical flow. At
higher $p_t$, all $v_2$/n measurements are very close in value
("constituent quark scaling") indicating coalescence of co-moving
constituent quarks. If coalescence is indeed the hadron production
mechanism, then it seems natural to conclude that a deconfined phase
of quarks and gluons is created prior to hadronisation. Note that
gluons do not seem to be present at hadronisation. The splitting
below $p_t$/n of 500 MeV/c appears to have mass dependence (a
signature of hydrodynamical flow). In addition, the pion
distribution at low $p_t$ is expected to be affected by resonance
decays.

\begin{figure}[htb]

\insertplot{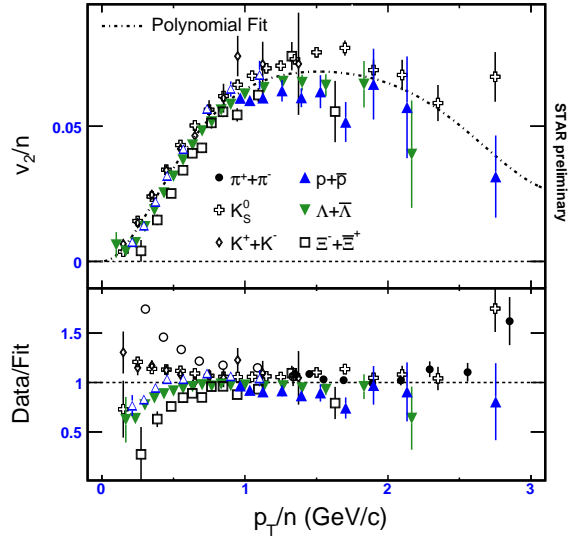} \caption[]{Measurements of scaled
$v_2$(p$_t$/n)/n for identified hadrons (upper panel) and ratio
(lower panel) between the measurements and polynomial fit through
all data points except pions for 200 GeV minimum bias Au+Au
collisions. Open symbols for $\pi^+$+$\pi^-$ and $p$+$\bar{p}$ were
taken from \cite{PH1}.} \label{fig4}
\end{figure}

To summarize: The quark-number dependence of $v_2$ suggests that the
relevant degrees-of-freedom are partonic, while the high degree of
collectivity developed by the even heavier strange quarks suggests
that the partonic degrees of freedom are locally equilibrated.

\section{Heavy Quarks as a Test of Early Thermalization}\label{techno}
Heavy quarks are created almost exclusively during first impact and
therefore are expected to address directly the early thermalization
of the system created in the collision. Unfortunately, at the
present time, direct measurements of charm and bottom at RHIC are
not yet feasible. Therefore $v_2$ of single electrons at
sufficiently high transverse momenta from non-photonic semi-leptonic
decays of heavy flavors is used as a substitute.

Preliminary analysis of experimental data by STAR and PHENIX show a
significant amount of flow for non-photonic electrons \cite{La,Ad},
which indicates flavor collectivity with heavy quarks flowing
together with light quarks.

While $v_2$ of non-photonic electrons clearly favors a non-zero
value at $p_t$ below 2 GeV/c, data obtained by different experiments
do not agree at higher $p_t$ ($2<p_t(e)<5 GeV/c$). These differences
are the subject of ongoing studies. If confirmed, heavy flavor
collectivity would imply light flavor thermalization.

\section{The Final Stage of the Collision - Jets and Thermalization}\label{techno}
The picture emerging from the evidence for equilibration and
near-ideal hydrodynamical flow is strongly supported by jet
quenching phenomena observed at RHIC. For the first time in heavy
ion collisions, the cross-section at RHIC is high enough for jet
production to play a measurable role. Due to their hard production
scales, jets, which materialize as a result of parton-parton
scattering processes very early in the collision, are embedded in
and propagate through the dense environment of the collision
"fireball" as it forms and evolves. Through their strong
interactions with the newly formed medium, partons lose energy,
before eventually fragmenting into ordinary hadrons, which preserve,
to a large degree, jet-like angular correlations. This energy loss
depends strongly on the properties of the evolving medium,
particularly on its energy density, and therefore the study of jet
quenching has become an important tool in the search for a
quark-gluon plasma at the earliest stage of the system evolution.

Experimentally, jets are selected by triggering on high $p_t$
particles, which are predominantly generated from a fast parton
escaping from the surface of the reaction volume. The other fast
parton created in the hard scattering is directed into the reaction
volume and traverses the medium. They are studied through the
angular correlations of associated fragmentation products of the two
jet partners. The early RHIC results clearly demonstrated nearly
complete disappearance of back-to-back correlations in central Au+Au
events (large path length in medium) and only small suppression of
the back-to-back correlation strength in peripheral collisions
(short path length in medium) \cite{QM}. The depleted jet energy
observed on the away-side (at $\Delta\phi \sim \pi$ from the
direction of the trigger particle) must be redistributed into low
$p_t$ particles. And indeed, it is the case as illustrated by the
left panel of Figure 5, where lowering the $p_t$ cut on the
away-side associated particles restores the missing away-side jet.
Reconstruction of these low $p_t$ particles is therefore fundamental
to understanding the fate of the energy lost by the primary parton
and provides a unique look at the details of the thermalization
process. Further study of associated particles has shown significant
differences in the spectral shape between p+p, d+Au and Au+Au
collisions. Figure 5 (left panel) presents the number and
$p_t$-weighted correlation functions in p+p, d+Au and central Au+Au
\cite{Ul,Fq} after background subtraction. The p+p and d+Au
distributions are similar, while the Au+Au is much broader.

\begin{figure}[htb]
\hspace*{0.1\textwidth}
\begin{minipage}{0.355\textwidth}
\psfig{file=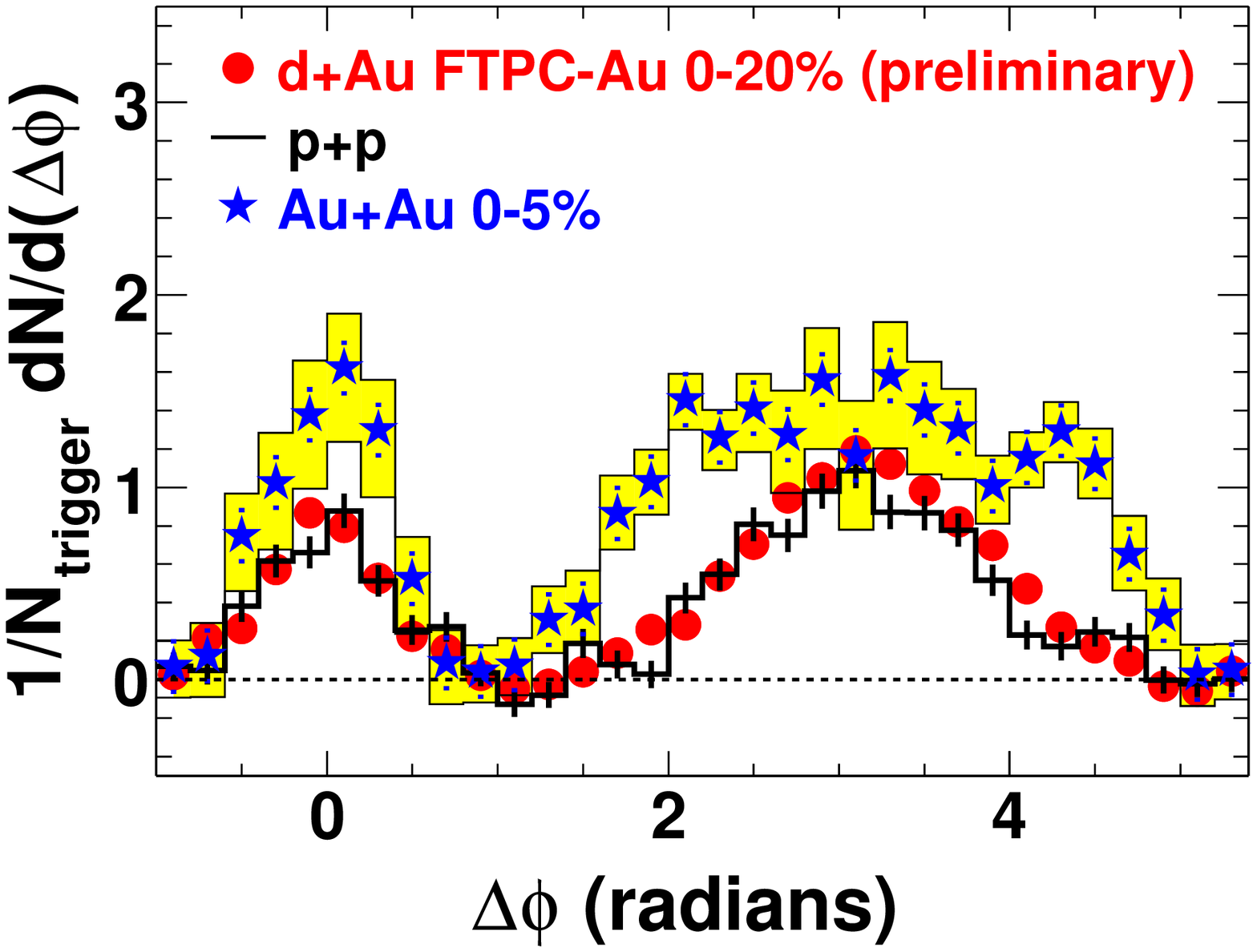,width=\textwidth} \vspace*{-0.34in}
\psfig{file=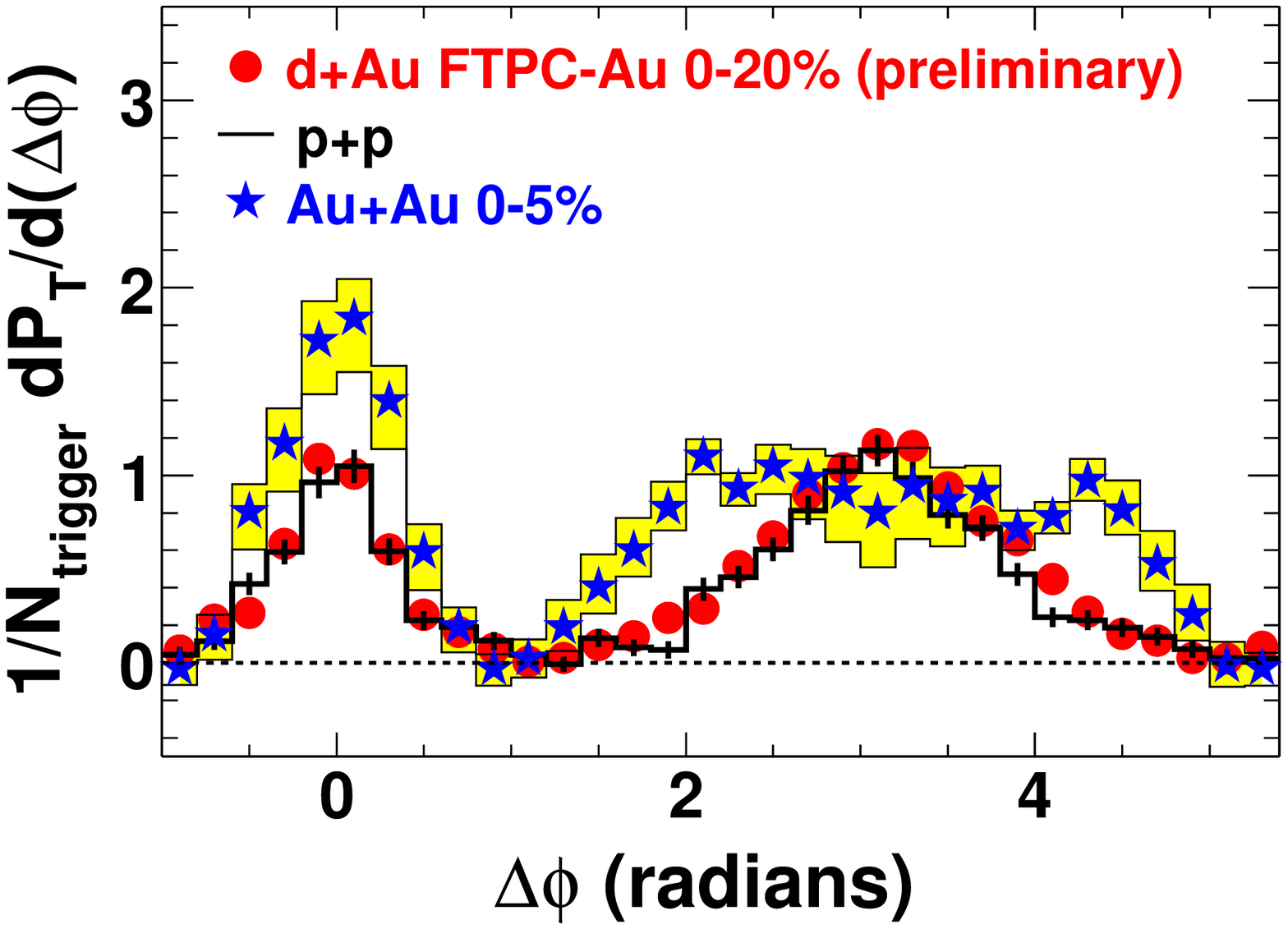,width=\textwidth}
\end{minipage}
\begin{minipage}{0.445\textwidth}
\vspace*{-0.2in} \psfig{file=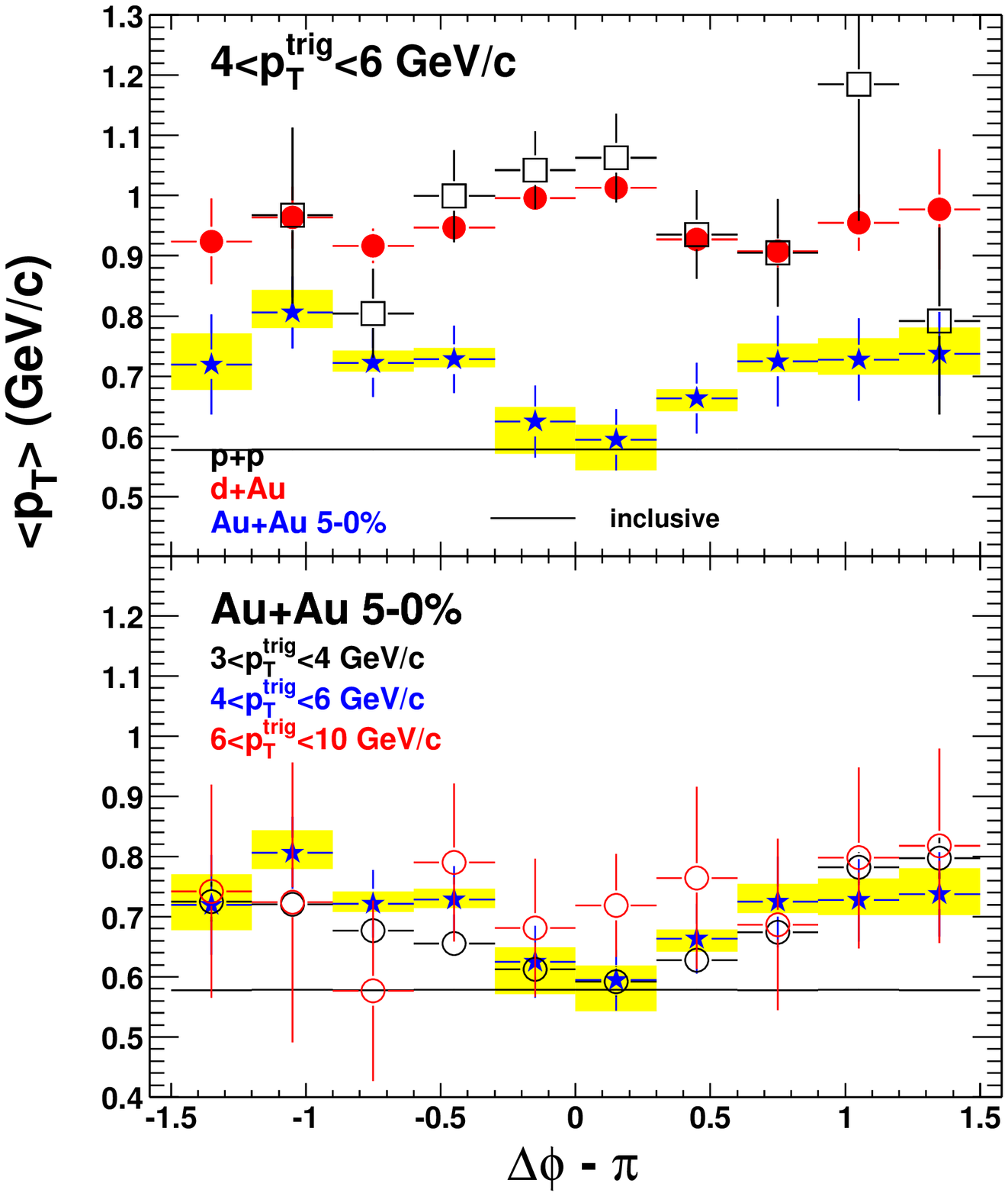,width=\textwidth}
\end{minipage}
\vspace*{-0.25in} \caption{Left: Background subtracted number (upper
panel) and $p_t$-weighted (lower panel) correlation functions in
$p+p$, central 20\% d+Au and 5\% Au+Au collisions~\cite{Ul}. Right:
The $\langle p_t \rangle$ of associated hadrons on the away side for
the three systems (upper panel) and three trigger $p_t$ selections
(lower panel). The shaded areas are systematic uncertainties.}
\label{fig2}
\end{figure}

Figure 5 (right panel) shows the $<p_t>$ obtained from the ratio of
the $p_t$-weighted and number correlation functions as a function of
$\Delta\phi$ on the away side. The $<p_t>$ for p+p and d+Au have
maxima at $\Delta\phi \sim \pi$, as expected from jet fragmentation,
while the $<p_t>$ for central Au+Au has a prominent dip at
$\Delta\phi \sim \pi$ and its value is similar to the inclusive
$<p_t>$ (marked by the straight line). The lower plot shows the same
$<p_t>$ behavior for the three trigger bins. The $<p_t>$ versus
$\Delta\phi$ results indicate that the spectrum in the direction
opposite to the trigger particle ($\Delta\phi \sim \pi$) is softer
than at different angles (away from $\Delta\phi \sim \pi$). This
means that fewer high $p_t$ particles survived the longer path
through the medium. This is also demonstrated in the correlation
functions with varying associated $p_t$ - see Figure 6. With
increasing associated $p_t$, the correlation function flattens and
even develops a double-peak structure \cite{Mr}. The two trigger
intervals presented (intermediate (left panel) and high $p_t$
triggers (right panel)) show very similar behavior. The new
structure of the away-side is beyond the experimental uncertainty
due to statistical (errors bars on Figure 6) and systematic
(histograms on Figure 6) errors. It is clear experimental evidence
of the bulk response to the energetic parton.

To summarize: The lower value of $<p_t>$ of associated particles at
$\Delta\phi \sim \pi$ indicates some degree of thermalization
established in Au+Au. A new shape of away-side associated
correlations is evident beyond experimental errors. It is clearly of
dynamical nature. A consistent picture describing the away-side
structures as a function of trigger and associated $p_t$ has yet to
be developed.

\begin{figure}[htb]
\hspace*{0.1\textwidth}
\begin{minipage}{0.355\textwidth}
\psfig{file=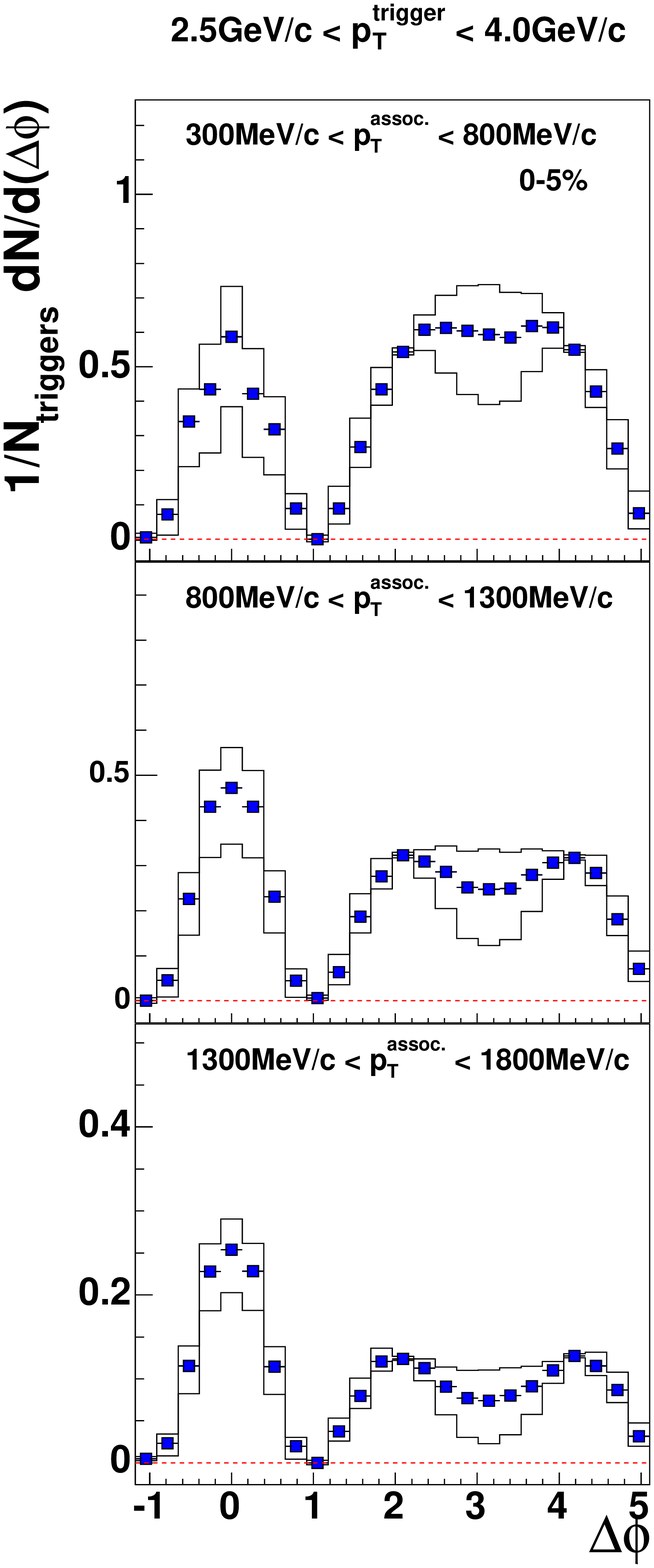,width=\textwidth}
\end{minipage}
\begin{minipage}{0.355\textwidth}
\psfig{file=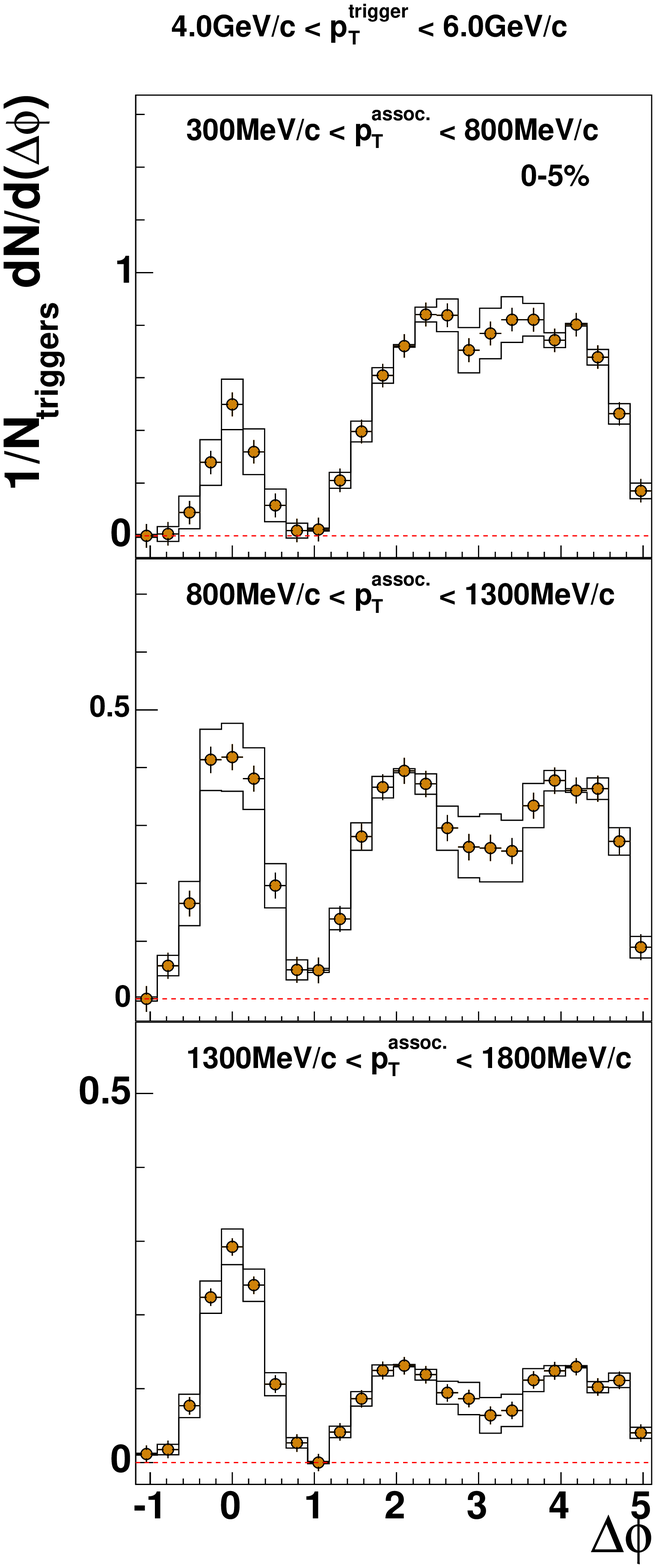,width=\textwidth}
\end{minipage}
\vspace*{-0.25in} \caption{Background subtracted number correlations
for two trigger bins, $2.5<p_t<4.0$ and $4<p_t<6.0$ GeV/c, for three
different associated $p_t$ windows. The evolution of the away-side
structure as function of $p_t$ associated is similar for two trigger
bins.} \label{fig2}
\end{figure}

\section{Three Particle Correlations}\label{techno}
The two particle correlation results are consistent with a number of
quite different scenarios. It is believed that analysis of three
particle correlations will allow the selection of the correct
theoretical description. Unfortunately, at the present moment,
results from different experiments are in disagreement. Studies of
the discrepancies are in progress.

\section{Conclusions}\label{concl}
The harvest of the first five years of data taking at RHIC is,
indeed, impressive. Pions, kaons, protons, electrons, and hyperons
have been measured in Au+Au collisions up to $p_t$$\sim$ 10 GeV/c.
Elliptic flow $v_2$ measurements made in Au+Au collisions revealed
collective behavior amongst partons (particularly important $\phi$
and $\Omega$ $v_2$). The non-zero value of $v_2$ for non-photonic
electrons indicates that interactions are copious enough for the u-,
d-, and s-quarks to be in a QGP state. Suppression of high $p_t$
particles, resulting from jet quenching, shows that the initial
density is high enough for partons to lose energy in partonic and/or
hadronic medium. The study of $<p_t>$ of the away side associated
particles demonstrates that even hard probes start to become
thermalized in the medium. Although the RHIC energy domain is not
fully explored yet, the progress in assessing the degree of
thermalization of Au+Au collisions (among other things) is
remarkable.

\section*{Acknowledgment(s)}
This work received support in part from the U.S. Department of
Energy  – Contract No. DE-AC02-05CH11231, Office of Nuclear Physics.

\section*{Notes}
\begin{notes}
\item[a]
Permanent address: Lawrence Berkeley National Laboratory,
Nuclear Science Division, Berkeley, CA 94720, USA;\\
E-mail: G\_Odyniec@lbl.gov
\end{notes}

\vfill\eject
\end{document}